\documentstyle[preprint,prb,aps]{revtex}
\begin{document}
\draft
\title{ Resonant Tunneling Spectroscopy of Interacting Localised States --
Observation Of The Correlated Current Through Two Impurities}
\author{V.V.Kuznetsov$^{(*)}$, A.K.Savchenko}
\address
{Department of Physics, Exeter University, Stocker Road, Exeter EX4
4QL, United Kingdom}
\author{D.R.Mace, E.H.Linfield, and D.A.Ritchie}
\address{Cavendish Laboratory, Madingley Road, Cambridge CB3 0HE,
 United Kingdom}
\date{\today}
\maketitle

\begin{abstract} We study effects of Coulomb interactions between localised
states in a potential barrier by measuring resonant-tunneling spectra with a
small bias applied along the barrier. In the ohmic  regime the conductance
of 0.2$\mu$m--gate lateral GaAs microstructures shows distinct peaks
associated with individual localised states.  However, when an electric
field  is applied new states start contributing to the current, which
becomes a correlated  electron flow through two interacting localised
states. Several situations of such correlations have been observed, and the
conclusions are confirmed by calculations.  \end{abstract}


\narrowtext
\tightenlines

Resonant--tunneling (RT) spectroscopy is a well known technique for the
study of  localised states in a potential barrier  \cite{fowler}. Its
essence is in  the measurements of the conductance in a transistor
microstructure where the barrier height is controlled by the gate voltage,
$V_g$. With changing  $V_g$ localised levels are moved with respect to the
Fermi level  in the contacts and a conductance resonance occurs when a state
passes the Fermi level.  It has been usually  assumed that there is no
Coulomb interaction  between the states, which is only justified if they are
well separated along the barrier width
 $W$, Fig.1a.
Otherwise the energy resonance of a state will depend on the occupancy of
the neighbouring states. When level $i$ is occupied, it will shift level $j$
upwards  by the Coulomb energy $U_{ij}\approx e^2/\kappa r_{ij}$, where
$r_{ij}$ is the separation between the two impurities and $\kappa$ is the
dielectric constant. We demonstrated that in GaAs structures with a doping
of $10^{17}$ $cm^{-3}$  and a gate length of 0.2$\mu$m there is a high
enough probability for several impurities to be close to each other and
interact  \cite{rrt}. These experiments were performed by measuring RT
spectra in the ohmic regime and suggested that, due to the  Coulomb shifts
of three interacting impurities, one level can exhibit two resonances in
$G(V_g)$.

In this work on the same type of structure we have found direct
manifestation  in the RT spectra of the Coulomb interaction between two
localised states -- a correlated current through two impurities. When a
small bias $V_{sd}$ is applied to the barrier, two interacting states can
only carry the current  in a correlated way: at the times when one level is
occupied the other level is lifted up and  switched off from conduction.

There have been intensive theoretical studies of tunneling through a
mesoscopic barrier containing interacting states \cite{RTint,glazman,raikh},
with several suggested models of the correlated current through two energy
levels.  In the situation where these levels were two  different--spin
states of the same impurity there was a prediction that the  averaged
conductance of a macroscopic barrier with many states should decrease with
magnetic field  \cite{glazman}. This effect was  observed in the hopping
magnetoconductance of large--area $\alpha$Si barriers \cite{beasley}.  For a
small--area barrier containing two impurities current--voltage
characteristics were calculated  for the case when electrons sequentially
tunnel through the two impurities  \cite{raikh}. Here we present
experimental results on mesoscopic GaAs barriers where the current through a
few localised states is measured as a function of $V_g$ at different biases
$V_{sd}$. This allows one to separate the contributions of the impurities to
the current. The results, confirmed by the  calculations, show that we are
dealing with several situations of the correlated current through two
interacting impurities which conduct in parallel.

Fig.1a shows the cross-section of our system, a metal--semiconductor
field effect transistor (MESFET), which is a
doped GaAs layer of $0.15\mu m$ thickness, MBE grown on an insulating substrate,
with a  short metal gate on the
surface of the wafer. Due to depletion under the gate,
a lateral barrier is formed in a doped layer and its height is
controlled by a negative voltage between the gate
and the 'source'--contact. Outside the gate area, conduction has metallic--like
character due to the high concentration of Si dopant ( $10^{17}$ $cm^{-3}$).
The differential
conductance between the source and drain was measured as a function of $V_g$,
using an AC lock-in technique, at different DC source-drain voltages, $V_{sd}$. The
measurements were performed in a dilution refrigerator where the temperature of
electrons in the sample was $\approx$100mK.

Consider an energy level which is moved downwards  with increasing $V_g$,
Fig.1b. With a source--drain bias applied to the barrier the level will show
two resonances in  the differential conductance $dI/dV_{sd}(V_g)$, namely at
$\mu_l$ and $\mu_r$ -- the energies where the current is switched 'on' and
'off'. The two resonances will form a cross-- like picture if
$dI/dV_{sd}(V_g)$--dependences are plotted with offsets for different
$V_{sd}$ (upwards for positive and downwards for negative biases), Fig.1b.
For both polarities of the bias, in the
 region to the left of the cross the state is positioned below  the
lowest of the two Fermi levels  and is occupied at zero temperature, and in
the region to the right of the cross it is empty. In the strip of width
$eV_{sd}$ between the two Fermi levels  the state is  positioned against
occupied states in one contact and empty states in the other and hence
conducts the current. In $dI/dV_{sd}(V_g)$ the conducting energy strip
corresponds  to the area between the lines of  the cross.

For non--interacting states one would expect a RT spectrum with a simple
superposition of their crosses which are shifted along the $V_g$--scale in
accordance with the  resonances of these states at $eV_{sd}=0$. However, the
experimental picture shown in Fig.2a is different.  Three resonances: at
$V_g=-1.8233V,\ -1.8215V$ and $-1.815V$ are seen in the ohmic regime. These
resonances correspond to RT through one  impurity since  a specific feature
of RT through two impurities \cite{2rt} is a strong  suppression of the peak
in $dI/dV_{sd}$ by an electric field with a negative  differential
conductance, which is not the case in the figure. As expected, the
resonances form the
 corresponding crosses when a $V_{sd}$ is applied.  In addition to the main
lines of the crosses, however, some extra lines are also seen in the figure.
Let us concentrate on the resonance at $V_g=-1.8215V$ and call the localised
state which gives rise to this resonance the 'first'  impurity. The
additional lines appearing in its cross manifest two other energy states
which start  contributing to the current only when a bias is applied.  A
characteristic feature of these extra lines is that they only exist $within$
the
 cross, when the first impurity carries the current. Extrapolation of these
lines shows  no resonance at their intercept with the $V_g$--axis which
means that in the ohmic regime,  when the first impurity is either empty or
occupied, these states have their  resonances elsewhere in $V_g$. Another
feature of the additional states needs an explanation: one state contributes
to the current primarily at a positive bias (at the top part of the graph)
 while the other is only seen at a negative bias.

Similar extra lines have been seen in single--quantum dot  structures and
interpreted as being due to the excitation levels of the quantum dot
\cite{weis}. In our case the extra lines are shifted with respect to the
main lines by $\Delta V_g\approx -0.82mV$  and $\Delta V_g\approx 0.72mV$.
This corresponds to $-205\mu eV$ and $180\mu eV$ in  the energy scale for
the rate of the
 Fermi level movement $\alpha=d\mu/deV_g\approx 0.2$, Ref.3.
This separation  is much smaller than one would expect for the separation
between the ground state and excited states of an isolated donor, $\Delta
\varepsilon \approx 4 meV$, and we  attribute the extra lines to two
separate impurities, 'second' and 'third', which are positioned close to the
first impurity.

When all three impurities are empty at small $V_g$ they are close to each
other in energy: level 3 is below level 1 by  $205\mu eV$ and level 2 is
above level 1 by $180\mu eV$, Fig.2b.  Consider how these energy levels are
affected in the ohmic regime when the impurities become charged. Suppose a
'normal' sequence of events, when impurities are charged  one by one and no
re--entrant resonance occurs when an occupied level is lifted up and becomes
empty again \cite{rrt}. When the Fermi level is raised and the third
(lowest) level becomes occupied
 it lifts levels 1 and 2 up by $U_{31}\approx e^2/\kappa r_{31}$ and
$U_{32}\approx e^2/\kappa r_{32}$, respectively. With $V_g$ increasing
further level 1 then  becomes occupied and lifts up level 2 by another
shift, $U_{12}\approx e^2/\kappa r_{12}$.  As a result in the ohmic regime
the  resonances due to the three levels are well separated in $V_g$, Fig.2b.
Indeed, for the interaction not to be screened by the metallic gate, the
distances $r_{ij}$ between the impurities have to be within $\approx 1000$
$\AA$, which is  the distance between the gate and the conducting channel.
This gives $U_{ij}\approx 1 meV$ and $\Delta V_g\approx 5 mV$ -- a much
larger shift than the separation between the main and the extra lines in
Fig.2a.

However, when a bias is applied and impurity 1 carriers the current, it is
only $partially$ empty and occupied. At the times when it
is occupied it lifts level 3 up and brings it back into the vicinity of
level 1.  Then level 3 comes into the conducting energy strip where it can
also carry the current. At other times, when level 1 is empty, level 2
comes down from its remote resonance and also
starts conducting. Therefore, in the area between the two lines of the
cross, levels 3 and 2 carry the current in correlation with the
occupancy of the the first impurity.

To explain why level 2 is only seen at a negative bias, let us
calculate the current through two interacting impurities, 1 and 2.
At $V_{sd}=0$ the shape of the $G(V_g)$ peaks in Fig.2a is well described
by a Lorentzian which is smeared by the Fermi distribution
in the contacts with an effective temperature of $\approx 100mK$ :
\begin{equation}
G(V_g)=\frac{e^2}{h}\frac{\Gamma_l\Gamma_r}{\Gamma_l+\Gamma_r}
\frac{sech^2((\varepsilon_i-\mu)/2k_BT)}{4k_BT}
\label{eq1}
\end{equation}
where $\Gamma_{l,r}\sim E_0exp(-2r_{l,r}/a)$ are the leak rates  from the
impurity to the left and right contacts respectively,  $r_{l,r}$ are the
distances from the impurity  to the contacts, $E_0$ is the ionisation
 energy and $a\approx 100\AA$ is the localisation radius.
For the peaks in Fig.2 the
smallest of the two $\Gamma$'s is $\approx 0.1\mu eV$ while the largest
is less than $10\mu eV$.
As the temperature smearing is smaller than the level separation,
$k_BT<<\Delta\varepsilon_{12}$,
 but larger than the width of the resonance, $k_BT>>\Gamma$, the kinetic
equations can be used to calculate the average occupancy of the two
impurities $\langle n_1\rangle$ and  $\langle n_2\rangle$, in a similar way
as is done for an impurity with two levels in Ref.4. For a large Coulomb
shift, $U_{12}>>\Delta\varepsilon_{12},|eV_{sd}|$,  the two impurities
cannot be occupied simultaneously and  $\langle n_1n_2\rangle =0$. Then:
$$\frac{d\langle n_1\rangle }{dt}=\Gamma_l^{(1)}\biggl(f_l^{(1)}\langle
(1-n_1)(1-n_2) \rangle -(1-f_l^{(1)})\langle n_1(1-n_2)\rangle  \biggr)$$
\begin{equation} +\Gamma_r^{(1)}\biggl(f_r^{(1)}\langle
(1-n_1)(1-n_2)\rangle -(1-f_r^{(1)})\langle n_1(1-n_2) \rangle \biggr),
\label{eq2} \end{equation} $$\frac{d\langle n_2\rangle
}{dt}=\Gamma_l^{(2)}\biggl(f_l^{(2)}\langle (1-n_1)(1-n_2) \rangle
-(1-f_l^{(2)})\langle n_2(1-n_1)\rangle\biggr)$$ \begin{equation}
+\Gamma_r^{(2)}\biggl(f_r^{(2)}\langle (1-n_1)(1-n_2)\rangle
-(1-f_r^{(2)})\langle n_2(1-n_1) \rangle \biggr), \label{eq3} \end{equation}
where $f_{l,r}^{(1,2)}$ are the distribution functions in the left and right
contacts at the levels $\varepsilon_1,\ \varepsilon_2$. Solving these
equations in  a steady state gives the expressions for $\langle
n_{1,2}\rangle$ in terms of the occupancies of the two states without
Coulomb interaction $\nu_{1,2}$: \begin{equation} \langle n_{1,2}\rangle
=\nu_{1,2}\frac{1-\nu_{2,1}}{1-\nu_{1,2}\nu_{2,1}}, \label{eq4}
\end{equation} where  \begin{equation}
\nu_{1,2}=\frac{\Gamma_l^{(1,2)}f_l^{(1,2)}+\Gamma_r^{(1,2)}f_r^{(1,2)}}
{\Gamma_l^{(1,2)}+\Gamma_r^{(1,2)}} .\label{eq5} \end{equation} The total
current through the two levels is given by
$$I=\frac{e}{h}\Gamma_l^{(1)}\biggl(f_l^{(1)}\langle (1-n_1)(1-n_2)\rangle
-(1-f_l^{(1)}) \langle n_1(1-n_2)\rangle\biggr)$$ \begin{equation}
+\frac{e}{h}\Gamma_l^{(2)}\biggl(f_l^{(2)}\langle (1-n_1)(1-n_2)\rangle
-(1-f_l^{(2)})\langle  n_2(1-n_1)\rangle\biggr), \label{eq6} \end{equation}
and after simplification \begin{equation} I=\frac{e}{h}\bigl(1-\langle
n_2\rangle \bigr)\frac{\Gamma_l^{(1)}\Gamma_r^{(1})}
{\Gamma_l^{(1)}+\Gamma_r^{(1)}}\bigl(f_l^{(1)}-f_r^{(1})\bigr)
+\frac{e}{h}\bigl(1-\langle n_1\rangle
\bigr)\frac{\Gamma_l^{(2)}\Gamma_r^{(2)}}
{\Gamma_l^{(2)}+\Gamma_r^{(2)}}\bigl(f_l^{(2)}-f_r^{(2)}\bigr) \label{eq7}
\end{equation}  The correlated current in Eq.(\ref{eq7}) is different from
the current through two non-- interacting impurities by the coefficients
$(1-\langle n_2\rangle )$ and $(1-\langle n_1\rangle )$. They reflect the
condition that one level can only carry the current if the other level is
empty. If the two impurities have a similar position  along the barrier
length, that is $\Gamma_{l,r}^{(1)}=\Gamma_{l,r}^{(2)}= \Gamma_{l,r}$, their
occupancies are also equal, $\nu_1=\nu_2=\nu$, and the current becomes
\begin{equation}
I=\frac{e}{h}\frac{2}{(1+\nu)}\frac{\Gamma_l\Gamma_r}{(\Gamma_l+\Gamma_r)},
\label{eq8} \end{equation}  As a result the value of the correlated current
is controlled by the occupancy $\nu$,  which depends on  the position of the
impurities along the barrier and also on the sign of the applied bias.
Suppose that the two impurities are positioned closer to the left contact,
Fig.3a, so that $\Gamma_l >> \Gamma_r$. Then $\nu\approx 1$ for a positive
bias
 from  Eq.(\ref{eq5}), as the states are then close to the contact with
occupied states  at energies  $\varepsilon _{1,2}$. The current through two
impurities in this case, Eq.(\ref{eq8}), is not  different from the current
through  only one impurity . Therefore, when with increasing $V_{sd}$  level
2 comes to the conducting strip it does not change the total current and no
extra line occurs in the differential conductance.  A positive  bias
corresponds  to strong correlations between the two impurities when
electrons on them have a long  'waiting' time before they leave to the right
contact.

For a negative bias electrons have a short waiting time, $\nu\approx 0$, as
the impurities are now close to the empty contact and the
correlations are weak. The current through two impurities is twice that for
level 1,  Eq.(\ref{eq8}), and  an extra line in $dI/dV_{sd}$ appears in the
first--impurity cross at a negative bias.

For correlations between the currents through impurities 1 and 3 in Fig.3b,
the situation is the inverse as level 1 has to be occupied for level 3 to be
brought into the conducting energy strip. For these impurities being closer
to the left contact an increase in the current  occurs when correlations are
strong, that is for a positive bias.

In Fig.4a another experimental result is presented, where for a negative bias the
appearance of a new state within the cross
is accompanied by the simultaneous suppression of a
main line of the cross and a region of negative differential conductance.
To explain such a decrease in the total current caused by level 2, consider a
situation when two correlated impurities carry
different currents. Suppose that level 2 is closer to the right contact,
$\Gamma_r^{(2)}>>\Gamma_l^{(2)}$,  while level 1 is close to the centre of
the barrier, $\Gamma_l^{(1)}=\Gamma_r^{(1)}$. The results of calculations
for such a situation, based on Eq.(7),
 are presented in Fig.4b and show a good agreement with
the data in Fig.4a.
The effect can be explained as follows.
The current through an impurity is limited by the
smallest of its leak rates $\Gamma_r$ and $\Gamma_l$, and it is  much
smaller for level 2 than  for level 1. This is why the total current  does
not change significantly when level 2 appears in the conducting strip at
positive biases when correlations are weak for this configuration of
impurities and no extra line is seen there.  However,  the low conducting level
can block the current through the highly
conductive level in the case of strong correlations at a negative $V_{sd}$.
Then level 2, which is closer to the filled contact, is occupied for a longer
time, $\nu_2>>\nu_1$, and  it lifts level 1 up and
decreases the total current.

In conclusion, we have presented the first demonstration of the correlated
conduction through two localised states in the parallel resonant--tunneling
channels. By studying RT spectra of localised states with a small bias along
the barrier, we have directly shown that two levels, which are separated by
a Coulomb shift when one is  fully occupied, can still conduct in a
correlated way.

We thank M.Raikh for reading the manuscript and valuable comments.

(*) Present address:  Department of Physics, State University of New York,
Stony Brook, NY 11794.

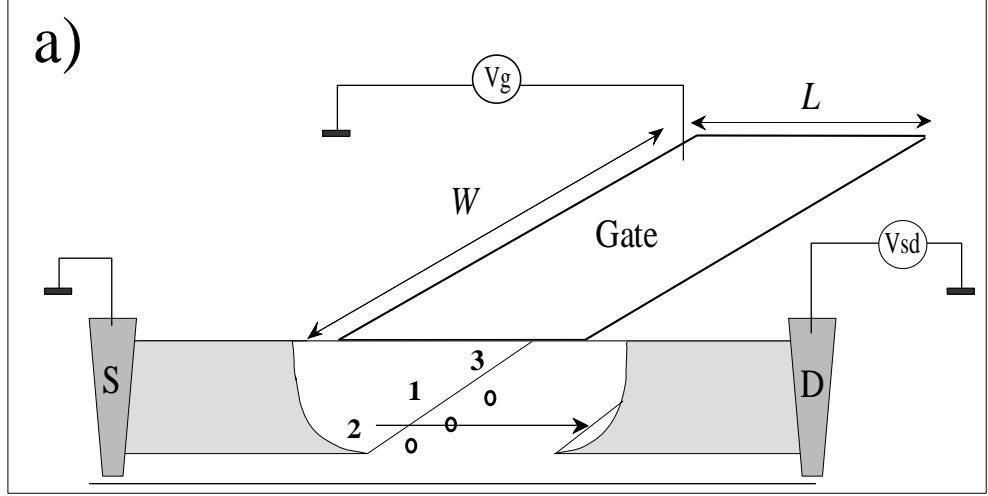
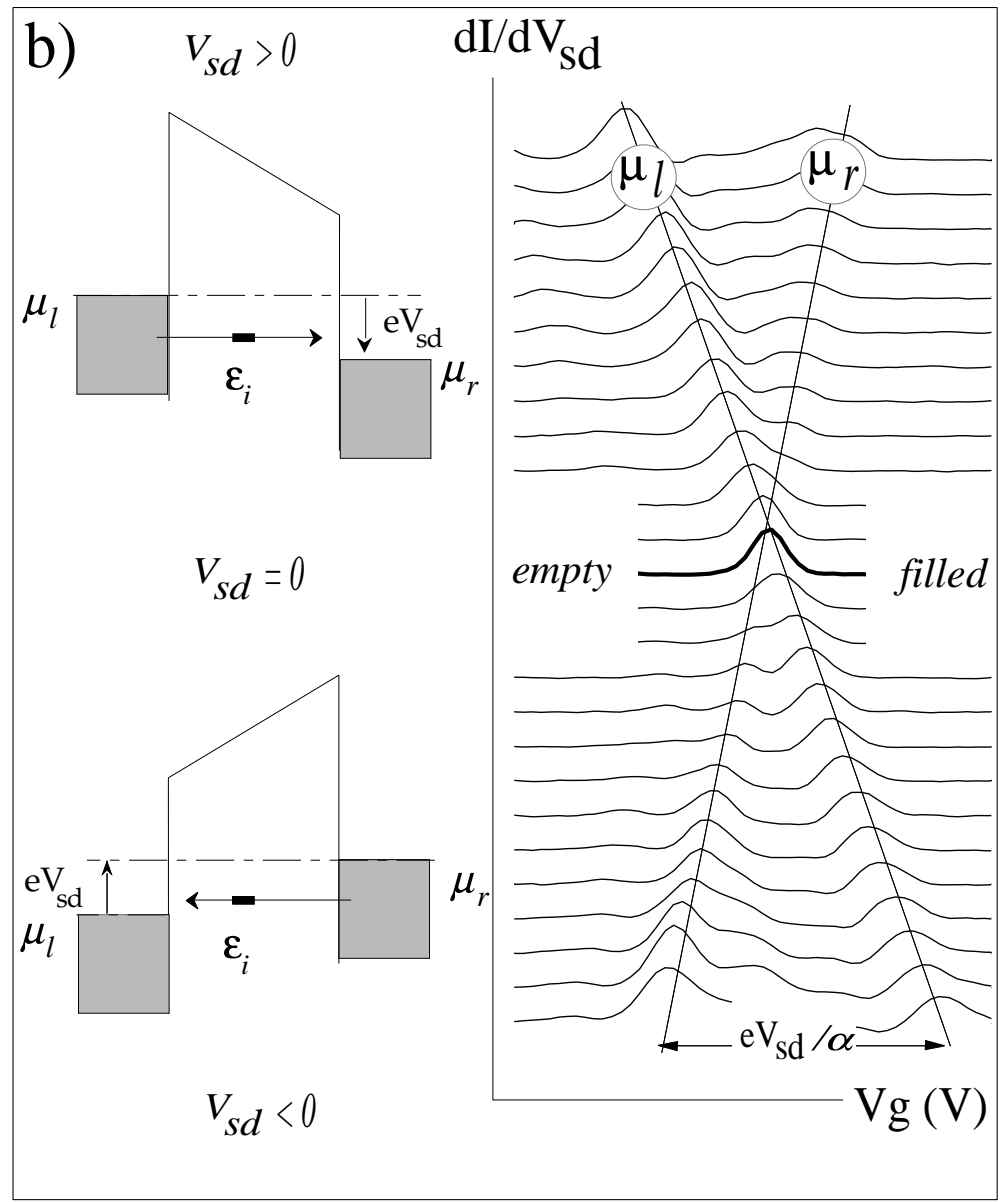
\begin{figure} \caption{a) Sketch of the geometry of the conducting channel
between two ohmic contacts in a GaAs MESFET, showing the position of a short barrier
and three impurities in it.
The gate length (along the current) $L$ equals $0.2\mu m$
and width $W$ equals $20 \mu m$.
b) Schematic diagram of the potential barrier with a positive and negative bias
applied. The
differential conductance as a function of gate voltage shows a 'cross' when
plotted with an offset for
 different $V_{sd}$ (upwards for a positive bias). The area between the two
resonances corresponds to the impurity carrying current and being partially
occupied.} \label{autonum}  \end{figure}

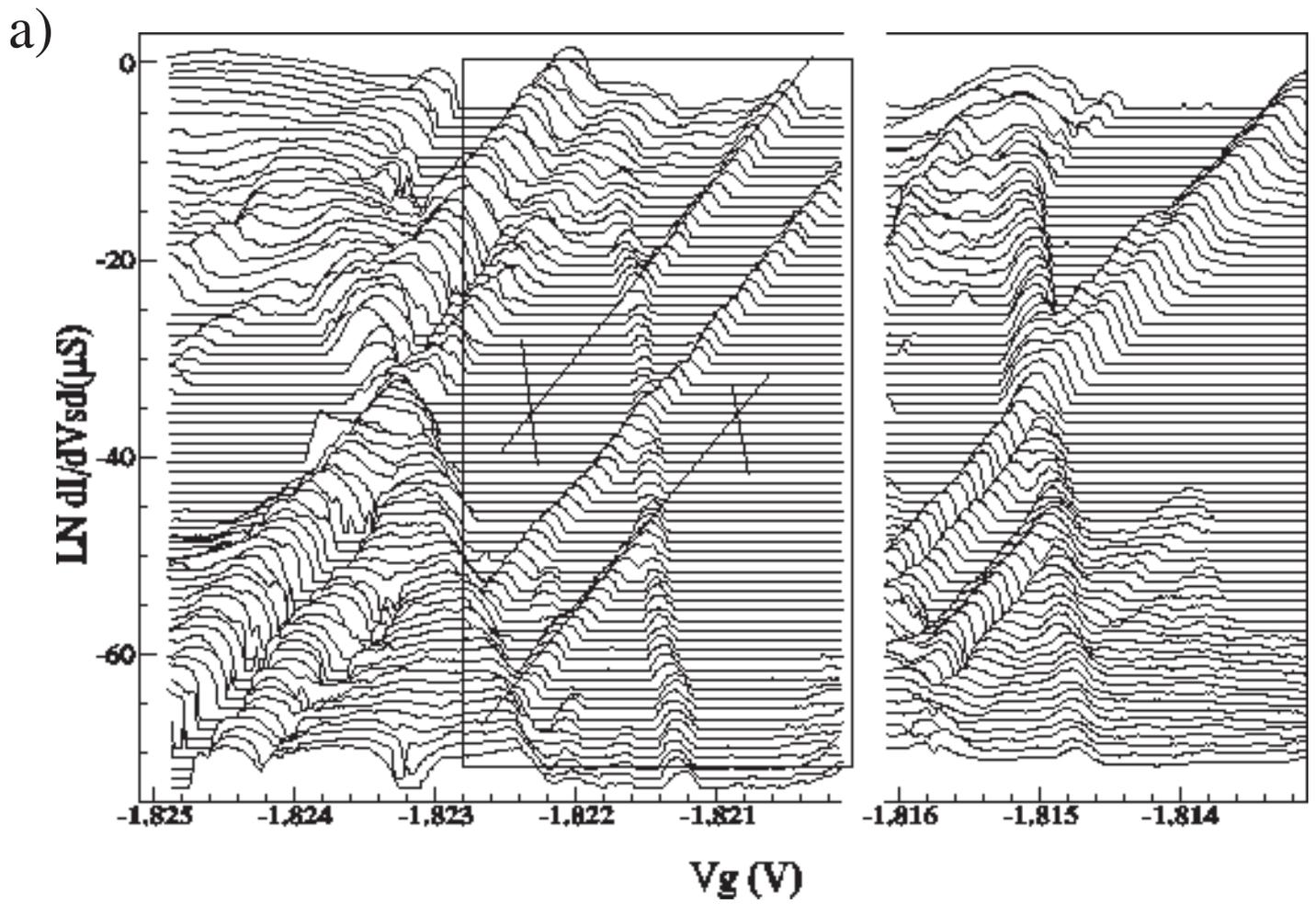
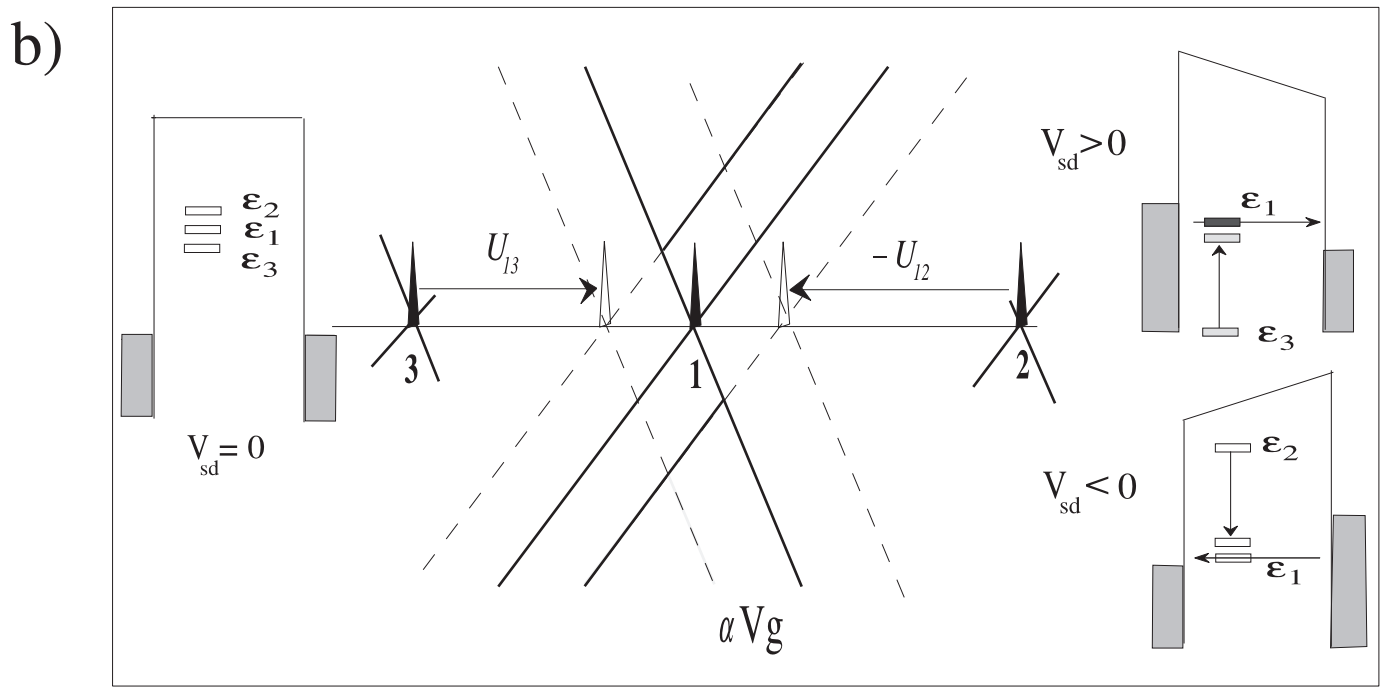
\begin{figure} \caption{a) Logarithm of the differential conductance as a
function of gate voltage at $T=100mK$ shown with a conductance threshold of
$0.01\mu S$.  Curves for different $V_{sd}$, changed with a step of $16.4\mu
V$, are offset ( upwards for a positive $V_{sd}$ with respect to the ohmic
regime shown by the highlighted curve).  b) Three interacting localised
states with close energies when not occupied. A diagram of the first
impurity cross with the extra lines within it due to impurities 2 and 3. In
the ohmic regime the resonances of these impurities are separated from the
first--impurity resonance, but with a $V_{sd}$ applied they are  brought
into its vicinity.}  \label{autonum}  \end{figure}

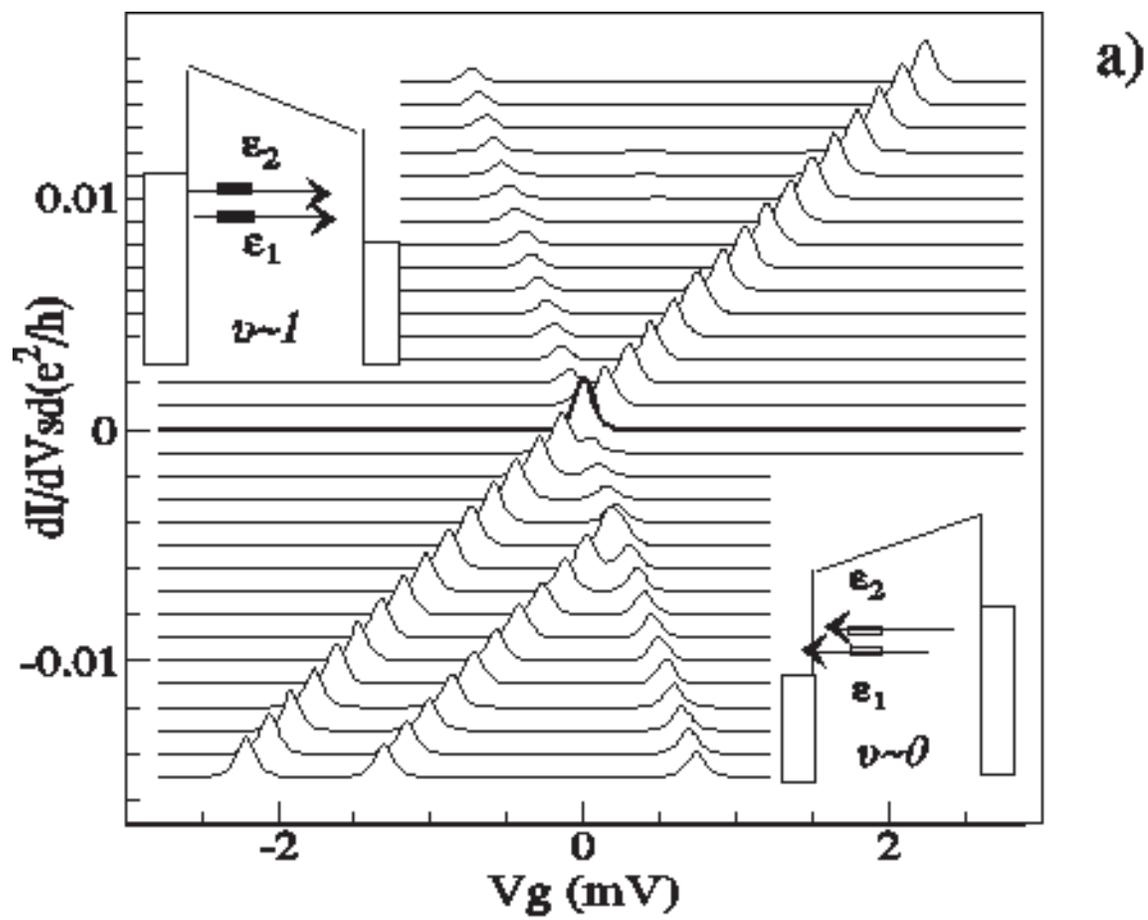
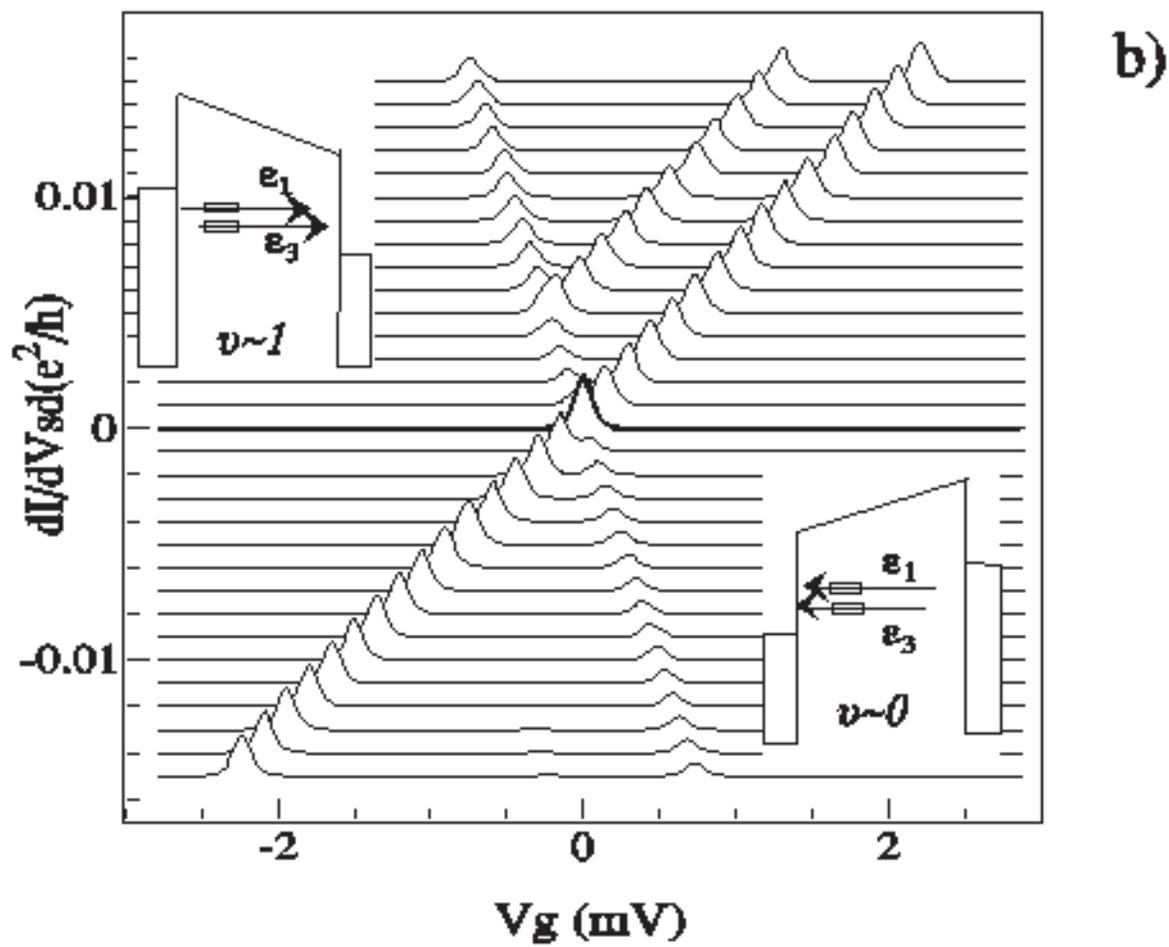
\begin{figure} \caption{ Calculations of the differential conductance for
correlated tunneling  through two impurities: a) 1 and 2,  b) 1 and 3. The
highlighted curves correspond to $V_{sd}=0$ and curves for different
$V_{sd}$, which is changed with  $\Delta V_{sd}=15\mu V$, are offset.
Parameters: $kT=10\mu eV$, $\Delta\varepsilon_{12}=\Delta\varepsilon_{13}=
230\mu eV$, $\Gamma_l^{(1)}= \Gamma_l^{(2)}=1\mu eV$,
$\Gamma_r^{(1)}=\Gamma_r^{(2)}=0.1\mu eV$,  $\beta=d\varepsilon_i
/deV_{sd}=0.25$, $\alpha=0.26$. Inserts show configurations for strong
($\nu=1$) and weak ($\nu=0$) correlations.}  \label{autonum}  \end{figure}

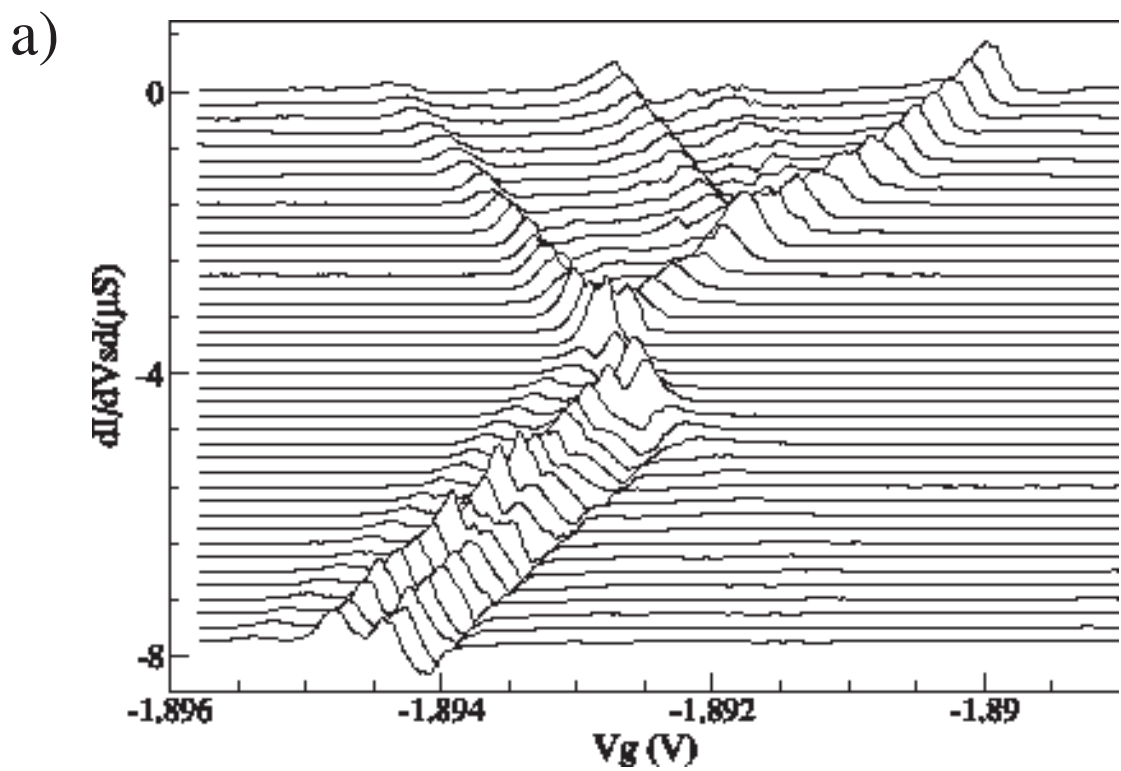
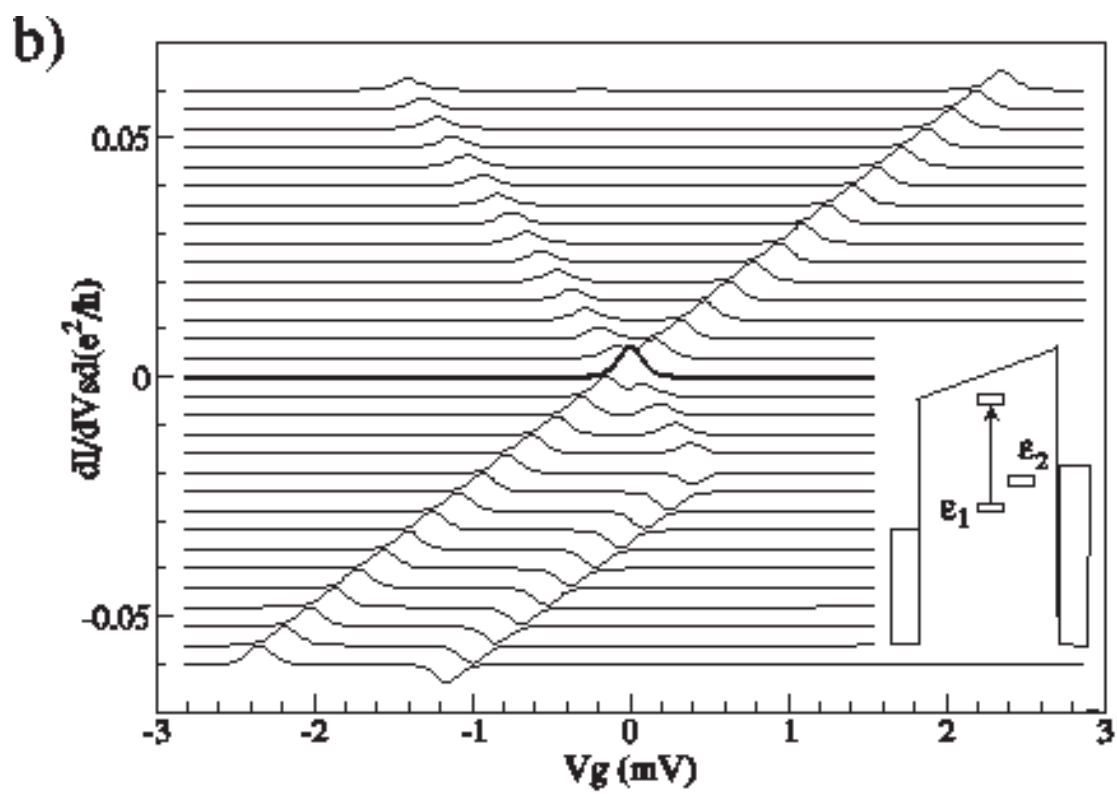
\begin{figure} \caption{a) Negative differential conductance occuring at a
negative $V_{sd}$, $T=100mK$.  The curves are measured with a step $\Delta
V_{sd}$ of  $50\mu V$.  b) Calculation of the differential conductance with
parameters: $kT=10\mu eV$,
$\Delta\varepsilon_{12}=230\mu eV$,
 $\Gamma_l^{(2)}=0.01\mu eV$, $\Gamma_r^{(2)}=5\mu eV$,
$\Gamma_l^{(1)}=\Gamma_r^{(1)}=0.5\mu eV$,  $\beta=d\varepsilon_i
/deV_{sd}=0.4$, $\alpha=0.2$, step $\Delta V_{sd}=50\mu V$.} \label{autonum}
\end{figure}

\end{document}